# Responsible AI Governance: A Response to UN Interim Report on Governing AI for Humanity

Open Consultation from the United Nations [(UN) AI Advisory Body](#).

## Executive Summary

**Opportunities and Enablers**

- We agree with the UN Interim Report: Governing AI for Humanity that AI has the potential to beneficially transform access to knowledge and increase efficiency in many spheres of life, in line with the Sustainable Development Goals (SDGs).
- At national levels, governments should foreground responsible, equitable, secure, reliable, and safe access to AI and inclusive input to include children, young people, vulnerable and marginalised people in the community. Regional policies should foster lawful and ethical data sharing practices and strengthen collaboration across borders.
- The networked nature of AI systems makes them deployed, hosted and used in different jurisdictions. Countries should regulate wherever systems are used locally, even if organisations are not incorporated locally.
- Investments in AI should include investments in infrastructure such as broadband and electricity, which support the use of AI systems.
- Different stakeholders should be aware of their responsibilities. For example, developers and engineers are responsible for their designs, developments and deployments (Yazdanpanah *et al.*, 2021; Deb *et al., 2023*), policy makers for holding players accountable (Urquhart, McGarry and Crabtree, 2022), and end users for their choices.

**Risks and Challenges**

- We agree that without appropriate safeguards, the use of AI systems may exacerbate inequalities in society. Further research needs to be conducted to investigate the claims for AI systems and, particularly in the education sector (Miao and Holmes, 2023) financial systems and critical infrastructure.

- Given the real potential harms of AI systems, the design, development, and deployment of AI systems should focus on the impact these systems have on the lives of all people, including those historically marginalised in society (Buolamwini, 2023).
- AI literacy will empower people to safeguard their privacy (Williams *et al.*, 2022), understand different options, and make informed decisions (Andries and Robertson, 2023).
- The environmental cost or impacts of AI systems should be more prominent in the report, particularly issues around regulations, governance, and the end of life of hardware and physical components of AI systems.

**International Governance of AI**

- There is a plethora of AI-related guidelines, frameworks, national strategies, standards, best practise documents, and codes (Jobin, Ienca and Vayena, 2019). Common themes amongst them are the need for inclusive and multi-stakeholder governance and the alignment of AI development with universal values, principles, and international norms.
- We see a challenge in integrating Equity, Diversity and Inclusion (EDI) perspectives in stakeholder interactions due to existing power asymmetries and biases inherent in the AI ecosystem.
- From a governance viewpoint, we suggest embracing non-Westphalian forms of governance, where states agree on common objectives, and digital governance, guidelines and recommendations may overlap but are not identical to digital regulation (Brownsword, 2022).

A. **Guiding Principles**
- To reduce the digital divide, AI tools should be made accessible, with a focus on vulnerable communities, which exist beyond the Global South and Global Majority.
- Digital divides should be viewed as plural, not singular – beyond access, we must consider logics, experiences, voices, ideas, well-being, and other intersections.
- Non-binding nudges should be supplemented with legally binding norms in AI governance, with less emphasis on self-regulation.
- Beyond data commons, other forms of data governance could be explored – van Geuns and Brandusescu (2020) suggest data cooperatives (i.e.,

collaborative pooling), fiduciaries (i.e., intermediaries between individuals and data collectors), trusts/stewards, sovereignty (i.e., shifting power to Indigenous Peoples and local people), marketplaces, intermediation (i.e., limiting data collected) and altruism (i.e., voluntary sharing for non-commercial purposes).

B. <u>Institutional Functions</u>
- We believe that transparency in decision-making processes is essential to fostering public trust.
- Professional bodies should provide clear guidelines on the use of AI systems in their respective areas of work. Standard Development Organisations (SDOs) like the International Standards Organisation (ISO), Institute of Electrical and Electronics Engineers (IEEE), European Telecommunications Standards Institute (ETSI), British Standards Institute (BSI), and National Institute of Standards and Technology (NIST) should and could lead this aspect.
- Addressing disparities in access to data, compute power, and talent requires innovative approaches and sustained investment in capacity-building and infrastructure development, particularly in regions with limited resources.

We thank the UN Secretary-General's AI Advisory Board for giving us an opportunity to respond to the Interim Report: Governing AI for Humanity.

# Suggested Citation

Sarah Kiden, Bernd Stahl, Beverley Townsend, Carsten Maple, Charles Vincent, Fraser Sampson, Geoff Gilbert, Helen Smith, Jayati Deshmukh, Jen Ross, Jennifer Wiliams, Jesus Martinez del Rincon, Justyna Lisinska, Karen O'Shea, Marjory Da Costa Abreu, Nelly Bencomo, Oishi Deb, Peter Winter, Phoebe Li, Philip Torr, Pin Lean Lau, Raquel Iniesta, Gopal Ramchurn, Sebastian Stein, and Vahid Yazdanpanah. Responsible AI Governance: A Response to UN Interim Report on Governing AI for Humanity, Public Policy, University of Southampton, 2024. DOI: https://doi.org/10.5258/SOTON/PP0057 **\***

**\***All the authors contributed equally.

**Suggested Citation Link in BibTex:**

eprints.soton.ac.uk/cgi/export/eprint/488908/BibTeX/soton-eprint-488908.bib

# How we arrived at this response

To identify contributors, an invitation was sent to Responsible AI (RAI) UK partners. Contributors to this response represented varying levels of experience within academia and industry, including senior academics, researchers, graduate students, and industry experts.

We engaged a range of academic practitioners in two activities: a survey and an online video call that was used for interactive brainstorming. In total, we had 13 survey responses and 20 attendees for the call. During the call on 22 February 2024, we used a visual interactive whiteboard to capture and discuss ideas for each of the areas identified in the Interim Report. Across these exercises, we sought to gather feedback and share output from our own research work as well as other relevant work in the field of AI that we believe supports the work that has been carried out by the UN Secretary-General's AI Advisory Board. Contributors volunteered as penholders and contributed to each of the sections, then worked iteratively to arrive at the response below.

# Our Response

## **Opportunities and Enablers**

We agree with the UN Interim Report: Governing AI for Humanity that AI has the potential to transform access to knowledge and increase efficiency in many spheres of life, in line with the Sustainable Development Goals (SDGs). For purposes of this response, we view actors in the AI ecosystem consisting of those researching, developing, deploying, working with, supporting, regulating or using AI systems (Jacobides, Brusoni and Candelon, 2021; The Alan Turing Institute, 2021). AI systems refer to different tools (both software and hardware), applications, models, and algorithms used to perform cognitive and computational tasks.

In response to the question to policymakers about growing successful AI ecosystems, we note that there are different levels of developing and implementing policies – at national level and regional and international levels. Taken together, these levels help to create an ecosystem that promotes safe and responsible AI for all stakeholders. At the national level, governments should foreground responsible, equitable, secure, reliable, and safe access to AI systems. Furthermore, they should create environments that foster risk mitigation, distribution of benefits and inclusive public and civic input to include children, young people, and vulnerable people in communities, to deploy and use AI systems that are beneficial for all peoples. This should be supported by investments in AI research and talent, innovative market structures, and incentives for companies that are contributing to AI for social good. Investments in AI should be supplemented with investments in infrastructure such as broadband, connectivity and electricity, which together support the sustainment of AI and other technical systems.

National strategies and policies form the basis on which AI systems can be deployed positively, while holding players accountable and responsible for their roles in the ecosystem. Thus, governance mechanisms should be clear and measurable. At regional and international levels, policies should foster ethical data sharing practices and strengthen collaboration across borders. One example of an organisation working to build data infrastructure is the Open Data Institute (ODI)[1], which is working with other organisations to create trust in data and data practices. Due to the networked nature of many technological systems (i.e., data can be stored in one location while the system is being used in another), collaborations will promote shared data repositories, data reuse and the extension of support to communities that are underserved in terms of access to technologies. We agree that, as highlighted in the Interim Report, there are some partnership examples to learn from; the Global Alliance for Vaccines and Immunisation (GAVI) Alliance[2], for instance, has aided the redistribution of vaccines and allowed low-income countries to acquire vaccines. Other examples include open science initiatives like the European Molecular Biology Lab (EMBL)[3], an intergovernmental organisation supported by 195 countries that is working on molecular biology research. We propose a similar distributed model for AI collaboration and see the UN as a good venue to facilitate such arrangements. Collaborations in the AI ecosystem can extend to international research consortia, social projects, and governance.

Cross-border AI projects should have clear SDG targets that are measurable. Notwithstanding, in cases of AI harms, countries should regulate wherever systems are used locally, whether the companies are registered locally or not. We need to think about cross-border data exchanges/transfers, especially in the context of countries that do not have, or have inadequate, data protection regulations or legislation.

Different stakeholders should be aware of their responsibilities. For instance, developers, designers and engineers who create products or services for commercial gain or for public end-users need to be aware that the algorithms and systems they design, develop and deploy have consequences, hence they should have responsibility for their code and machine outcomes (Yazdanpanah *et al.*, 2021; Deb *et al.*, 2023). Likewise, policy makers should be responsible for holding players accountable through robust regulatory frameworks that consider legally mandated design requirements for AI systems (Urquhart, McGarry and Crabtree, 2022). Different stakeholders have different levels of responsibility. For example, children using AI technologies in schools should not be held accountable if their choices were not clear. So, we should advocate for AI systems to be explainable, robust, reliable, secure, and trustworthy.

---

[1] Open Data Institute (ODI) https://theodi.org
[2] GAVI Alliance https://www.who.int/europe/initiatives/gavi-alliance
[3] European Molecular Biology Lab https://www.embl.org

It is a collective role; however, we need to ensure that there is capacity-building for all stakeholders. End users particularly will benefit from getting a basic understanding of ethics so as to make informed choices/decisions. Big technology corporations should collaborate with local communities, train, and fine-tune models on local data. We discuss this further in the section on Institutional Functions.

While this section of the Interim Report highlights governance as a key enabler, we believe that AI governance can only be achieved if other stakeholders and sectors in the AI lifecycle play their part. Correspondingly, there is a need to continuously monitor and analyse AI data and AI models to identify patterns and embed AI accountability into the ecosystem. We should recognise that getting universal buy-in on AI governance may take several years, if it is ever achieved, with several iterations of guidelines and norms.

## Risks and Challenges

AI systems may pose physical or psychological harms as a result of bias, errors, or malicious intent and those who deploy such systems to public end-users not respecting human rights. We agree that without effective safeguards, the use of AI systems may exacerbate inequalities in our society. It is important to think critically about applications of AI in facial recognition, policing, finance and insurance, and autonomous systems in order to reduce the potential for bias and misuse as these are critical sectors where AI inequalities are already being observed. Having access to data and resources is crucial for identifying risks in AI systems and allocating responsibility. Opacity in AI systems can be reduced by having explainable systems (The Royal Society, 2019) and opening proprietary datasets, models and systems for evaluation and auditing. However, we should consider different forms of accessing and using data. We discuss different alternatives to data governance below in our response for Guiding Principle 3, which include data commons, cooperatives, fiduciaries, and trusts. We believe that openness and transparency go hand in hand, ensure accountability, and lead to public trust in the long term.

With the increased reliance on AI systems, care should be taken to ensure that there are alternative options available when there is service downtime on a system that many people rely on. Some stakeholders are making bold claims about AI performance, yet further research needs to be conducted to investigate the claims for AI systems and, particularly in the education sector (Miao and Holmes, 2023), financial systems and critical infrastructure. Sharing findings from their research on how children interact with or understand AI, Andries and Robertson (2023) make a case for AI literacy, emphasising that literacy will safeguard children's privacy and help them make informed decisions. In another study on privacy in speech and audio, Williams et al. (2022) argue that privacy assurance and enforcement are still lacking even in cases where privacy is regulated or protected by law. They propose that individuals and end users should be able to decide their privacy levels based on informed choices. These two examples, among others, reinforce the need to build systems in a way that end users understand their options and can make informed decisions to meet their unique needs as they choose. It should be noted, though, that understanding how a system works does not necessarily give people the power to opt out. For instance, some privacy risks in systems like facial recognition in public spaces cannot be avoided. In such a case, the responsibility would not be put on end users. We believe that literacy programmes should target different types of end users. In schools, there should be wide scale programmes of education from as early as primary school.

Additionally, the tendency for techno-solutionism must be approached with caution because AI systems alone cannot solve all critical pressing problems. In broader research on Human-Computer Interaction (HCI), researchers like Grimes and Harper (2008) have advocated against technologies that are designed to address 'problems', and instead design celebratory technologies that work in tandem with existing human interactions, thereby keeping the human in-the-loop.

In their article on trustworthy AI as a framework, Saif and Ammanath (2020) suggest that businesses face challenges that are more human-based than machine-based. These challenges can be traced to human rights, ethics, and governance. Therefore, human agency and oversight become an integral part of developing trustworthy AI systems (Floridi, 2018). At the same time, it should be recognised that AI is increasingly embedded in complex socio-technical systems, where complex and highly distributed human-AI partnerships offer new opportunities but can also pose new emergent risks (Ramchurn, Stein and Jennings, 2021).

People are more likely to trust AI systems which are built ethically, safely and with consideration for humans in the system. As Buolamwini (2023) inquires, "*given the real harms of AI, how can we centre the lives of everyday people, and especially those at the margins, when we consider the design and deployment of AI*?". Similarly, Stein and Yazdanpanah (2023) advocate for citizen-centric AI systems, which are designed to benefit and closely involve human end users in their design, development and deployment. Through their work, the UN Advisory Body should emphasise human well-being and human-centric design to enhance user experience and promote public trust. Likewise, human knowledge should be valued and treated at the same level as machine-generated knowledge, if not greater. There is an emerging field of computational participatory modelling (Quimby and Beresford, 2022), which centres on knowledge co-production and the gathering of cross-sectional and longitudinal data. Such participatory and human-centred methods will move us towards trustworthy AI systems.

In their study focusing on migration and border governance, McGregor and Molnar (2023) call attention to challenges like discrimination, racism and human rights violations that are experienced by people at border points. Some of these instances include the improper use of identification tools (like facial recognition) or inferential tools (like lie detectors) and algorithmic risk assessments for visa sorting and decision-making for people at border points. While biometric person recognition has been viewed as an essential tool for refugees and asylum seekers, offering an identity verification method when passports may have been lost, the potential for such an AI system to misidentify vulnerable populations of people or violate their data privacy cannot be understated (Farraj, 2011).

We agree with authors like McGregor, Murray and Ng (2019), who contend that International Human Rights Law (IHRL) provides a framework with a shared understanding and means of assessing harm and assigning responsibility to different actors. They reiterate that appropriate checks and balances are needed to ensure that algorithms contribute to society while safeguarding against risks. A recent report by the Internet Watch Foundation (IWF) found that AI was being used to create child sexual abuse material (CSAM), with over 20,000 images posted in a one-month period (Internet Watch Foundation, 2023). Fundamental rights for the protection of children should be extended to the digital realm. Besides, there are other aspects that can be related to International Human Rights Law, comprising non-discrimination, even if not enforceable. These safeguards should apply to the full AI lifecycle.

We also note that while acknowledged, issues around the environmental cost of AI are not prominent in the report. Regulations and governance around this may be helpful to facilitate transparent reporting by companies and thinking about the end of life of the hardware/physical components of AI systems (including computers, peripherals, and other connected devices). In terms of the electricity demands of AI systems, it may be useful to look beyond pure energy consumption and consider the source of this energy, i.e., whether it is produced by clean renewable sources (de Vries, 2023).

## **International Governance of AI**

There is already a plethora of AI-related guidelines, frameworks, best practice documents and codes (Jobin, Ienca and Vayena, 2019). For example, in Europe, we have, among other things, the EU AI Act[4], the Council of Europe AI Convention[5] and UK's Pro-innovation approach to AI regulation[6]. Common themes amongst these documents are the need for inclusive and multi-stakeholder governance. Conversely, the frameworks take different approaches. For instance, the EU AI Act takes a centralised approach to AI governance, whereas the UK takes a risk-centred approach to AI regulation, making the implementation of international AI regulations complex. The importance of aligning AI development and deployment with universal values, principles, and accepted international norms also seems to be a common factor. The **implementation** of unified AI principles remains a major challenge, especially in integrating Equity, Diversity, and Inclusion (EDI) perspectives. Particularly, the **inclusivity** of minority groups, such as racial and ethnic minorities, civil society, and consumer/user groups. Stakeholder engagement has been primarily dominated by representatives from the industry and big technology companies, side-lining broader community voices. We need practical mechanisms to improve the EDI aspects of all stakeholders' participation. This means acknowledging and taking action to tackle systemic inequities, biases and power imbalances that are inherent in the AI ecosystem.

Most crucially, given the global nature of Responsible AI challenges, it has been recognised that international cooperation is paramount. Broad governance also needs to ensure the **interoperability** of different regulatory approaches and regimes, focusing on the common grounds for improving such interoperability. No unique and/or adaptive measure of governance can be complete without concerted international effort and political will. For holistic and effective governance, principles of AI and data governance should be harmonised internationally and in international trade agreements. We can see that governments' regulatory powers for AI security and auditing have been narrowed down on key issues including data localisation, data sharing, and the ban of disclosure of source code. Such AI and data governance clauses have widely emerged in regional trade agreements, which have negative impacts on AI **transparency**.

---

[4] The EU AI Act https://digital-strategy.ec.europa.eu/en/policies/regulatory-framework-ai
[5] The Council of Europe AI Convention https://www.coe.int/en/web/artificial-intelligence/cai
[6] AI regulation: a pro-innovation approach
https://www.gov.uk/government/publications/ai-regulation-a-pro-innovation-approach

What has also been missing, which has been partially addressed by the report, is the specific mention of the role of AI in addressing societal challenges such as climate change, public health, economic development, and crisis response through the development of public data commons in the form of open knowledge repositories. The frameworks discussed in the provided excerpts primarily focus on AI governance, standards, and risk management. While they emphasise the importance of addressing challenges and opportunities related to AI, there is indeed a lack of explicit mention of related technologies such as spatial computing, digital twins, and quantum computing (Kop *et al.*, 2023; Reichental, 2023). Spatial computing involves the use of digital technology to interact with the physical world in a spatial manner, often through Augmented Reality (AR) and Virtual Reality (VR) applications. Quantum computing, on the other hand, leverages principles of quantum mechanics to perform computations that traditional computers struggle with.

Given the rapid advancements in these technologies and their potential impact on society, it would be beneficial for governance frameworks to consider their implications alongside AI. Incorporating discussions on the Internet of Things (IoT), spatial and quantum computing into AI governance frameworks can help ensure a comprehensive approach to addressing emerging technological challenges and opportunities.

Moreover, we need to think about how we humanise AI and its implications. Anthropomorphism, the attribution of human-like characteristics to non-human entities, is increasingly prevalent in discussions surrounding AI, where machines are often depicted as possessing human-like intelligence, emotions, and intentions. For example, studies show that AI agents are becoming increasingly human-like in their physical appearance and ability to replicate emotions and display personality traits (Epley, 2018; Zhou *et al.*, 2019). This can shape public attitudes and expectations, as well as interactions with AI systems, potentially leading to misunderstandings and/or biases. Also, it brings up big ethical questions about how we treat AI and what rights an AI system may have, if any. Therefore, governance frameworks should also account for anthropomorphism to promote informed decision-making, mitigate risks, and safeguard human values and dignity.

From a governance viewpoint, one suggestion is to embrace a non-Westphalian form of governance where states agree on common objectives (Brownsword, 2022). This comes to mind:

> *"Digital governance may comprise guidelines and recommendations that overlap with, but are not identical to, digital regulation......Not every aspect of digital regulation is a matter of digital governance and not every aspect of digital governance is a matter of digital regulation" (Floridi, 2018).*

Re-inventing the wheel by incorporating new legislation is far from desirable - what might be useful is to progress adaptive governance mechanisms, include civil society organisations, and key population groups (those with lived experience) to form governance measures and mechanisms (Lau, van Kolfschooten and van Oirschot, 2023). For this, we might look towards other UN-level international governance that has been successful (e.g., using a human rights, decolonisation, or feminist lens as approaches) to foster international collaborative governance efforts.

## A. **Guiding Principles**

The Guiding Principles described in the Interim Report represent a holistic approach to AI governance with an emphasis on inclusivity, the public interest, data governance, multi-stakeholder collaboration and the alignment with existing laws and norms.

**Guiding Principle 1: AI should be governed inclusively, by and for the benefit of all**

AI opportunities, challenges and risks manifest globally for served and underserved individuals and communities. To reduce the digital divide, AI tools should be made accessible, with a focus on vulnerable communities, which exist beyond the Global South and Global Majority. Besides, digital divides should be viewed as plural and not singular – different access, logics, experiences, voices, and ideas. Ragnedda, Ruiu and Addeo (2022) categorise a digital divide as access to the Internet, competence to use digital technologies and personal wellbeing. We need to ensure that as we deploy AI systems, we are not creating situations where those without access to basic technologies are further marginalised (Gillwald, 2017).

**Guiding Principle 2: AI must be governed in the public interest**

Whereas non-binding nudges have their role in the ecosystem, we believe that these should be supplemented with legally binding norms so that players are held accountable for their actions. Additionally, relying purely on self-regulation should be discouraged as much as possible because it creates room for actors to opt for simpler options that are inexpensive and require little or no accountability. A recent example in the UK is the Post Office (Horizon System) scandal, where over 900 sub-postmasters and postmistresses were wrongfully convicted between 1996 and 2018 for loss of money caused by faulty computer software (BBC News, 2024). Decades later, the UK Government has responded by developing new legislation (the Offences Bill) to overturn the convictions and compensate people who were convicted using evidence from the Horizon system (Torrance *et al.*, 2024). Some of the victims of this scandal have passed away before receiving remedy for this injustice, further highlighting that enduring legacy that AI systems can have on society. There should be strict control over how authorities use such technology. The Justice and Home Affairs Committee of the House of Lords recommended a mandatory register of algorithms, introducing a duty of candour on the police, and setting up a national body which would develop strict scientific validity and quality standards (House of Lords, 2022). Incentives can be offered to actors who design, develop, deploy, and monitor AI systems in the public interest.

When advocating for representation of diverse stakeholders, we should acknowledge that power asymmetries occur in AI or technical ecosystems that are not guided by protective principles. What are the roles and responsibilities for each of the stakeholders? Who has the power to make AI design decisions?

**Guiding Principle 3: AI governance should be built in-step with data governance and the promotion of data commons**

The Interim Report stresses the value of data in major AI systems. Data anonymisation techniques and privacy-preserving technologies should be promoted so that data is anonymised and AI models are compliant before they are made public as data commons. In addition to the proposal for a data commons, other data governance models should be evaluated. In looking for the meaning of shifting power through data governance, van Geuns and Brandusescu (2020) identified seven alternative data governance models that include data cooperatives (collaborative pooling of data), data fiduciaries (intermediaries between individuals and data collectors), data trust (stewards), indigenous data sovereignty (shifting power and control from governments and institutions to Indigenous Peoples), data marketplace (if end users opt to sell their data for services or other benefits), data intermediation (limiting the amount and purpose of data collected), and data altruism (voluntary sharing of data for non-commercial purposes). These alternative models will allow individuals and communities to choose a model that works best for their lived contexts and the consequences of opting in or out of a particular model.

**Guiding Principle 5: AI governance should be anchored in the UN Charter, International Human Rights Law, and other agreed international commitments such as the Sustainable Development Goals**

We agree with the proposal for AI governance to be anchored in existing laws and norms. We offer that there is an urgent need to offer guidance to different categories of stakeholders on expectations, obligations, rules, safety and when to decline the use of a system. Smith, Downer and Ives (2023) note that it is unfair to ask people to use AI without proper guidance. To start with, procurement teams within police and other law enforcement offices need to receive guidance on the procurement of AI systems and other technology systems. There are several examples of challenges with the use of facial recognition software in policing. These include the Hikvision and Xinjiang example in the UK (House of Commons, 2021; Sampson, 2023) and other cases in the US (Buolamwini and Gebru, 2018; Buolamwini, 2023). In other sectors like healthcare, high-level guidelines around professional conduct could be transposed to inform the use of AI in the sector (Smith, Downer and Ives, 2023). Professional bodies should provide clear guidelines on the use of AI systems in their respective areas of work. Standard Development Organisations (SDOs) like the International Standards Organisation (ISO), Institute of Electrical and Electronics Engineers (IEEE), European Telecommunications Standards Institute (ETSI), British Standards Institute (BSI), and National Institute of Standards and Technology (NIST) should and could lead processes of developing guidelines. We recognise that standards developments are slow and getting consensus is challenging but believe that the results are rewarding for society. Beyond the suggested laws and norms, other international humanitarian law should be considered, including refugee protection.

As explained in the previous section on the governance of AI, there is an inordinate quantity of guidelines and principles, which may be daunting for small and medium enterprises (SMEs) looking to explore AI for business insight. How these guidelines are interpreted and applied is of utmost importance.

## B. Institutional Functions

The Interim Report outlines a comprehensive set of functions essential for governance of AI. While these functions provide a solid foundation, several critical issues and considerations arise:

**Institutional Function 1: Horizon scanning and building scientific consensus**

Establishing an independent, multidisciplinary body for AI assessment, like the Intergovernmental Panel on Climate Change (IPCC)[7] is crucial. However, ensuring its effectiveness requires addressing potential biases, conflicts of interest, and the influence of vested stakeholders. Transparency in decision-making processes and mechanisms to mitigate undue influence are essential to maintain credibility and foster public trust.

**Institutional Function 2: Interoperability and alignment with norms**

While a Global AI Governance Framework grounded in international norms is proposed, achieving global consensus on norms and standards can be challenging. Diverse cultural, political, and socio-economic differences among nations may make it complex to align on issues. Therefore, effective mechanisms for negotiation, compromise, and enforcement are necessary to overcome these barriers and ensure effective governance. Moreover, legally binding norms and enforcement need to be developed and considered at local/state level before being considered at global level such as the UN. This is because available remedies are critical to confidence (public and commercial), hence there is a need for legal systems that are prepared for enforceability. Transparency and explainability are key aspects of developing legal standards, as such evaluations and decisions must be interpretable to non-experts. Smith and Fotheringham (2020) discuss an example of medical negligence in the context of AI and healthcare. They describe different forms of responsibility for key stakeholders and call for a proactive approach to ensure patient safety. Clear guidelines, grounded in international norms like the Universal Declaration of Human Rights (UDHR)[8], will help to ensure that AI governance arrangements are interoperable across jurisdictions and all stakeholders are aware of their responsibility in the AI ecosystem.

**Institutional Function 3: Mediating standards, safety, and risk management frameworks**

Harmonising technical and normative standards across jurisdictions is vital. However, achieving consensus on complex technical issues and balancing the interests of various stakeholders, including industry, academia, and civil society, can be daunting. Transparent and inclusive processes for standard-setting are essential to address concerns regarding bias, favouritism, and exclusion. This includes worldwide bias, for example, adopting standards from a jurisdiction in the Global North to re-apply to the Global South where cultural needs may differ must also be considered. Therefore, it is imperative that global stakeholders have a voice in developing such frameworks.

---

[7] Intergovernmental Panel on Climate Change https://www.ipcc.ch
[8] Universal Declaration of Human Rights
https://www.un.org/en/about-us/universal-declaration-of-human-rights

**Institutional Function 4: Facilitation of development and use-liability regimes**

While international collaboration is crucial for AI development and deployment, ensuring equitable access to resources and benefits is paramount. Addressing disparities in access to data, compute power, and talent requires innovative approaches and sustained investment in capacity-building and infrastructure development, particularly in regions with limited resources. As underscored in the section on risks and challenges, transparent processes are required not just for digital border technologies but across AI systems used across national borders (McGregor and Molnar, 2023). Equitable access may be complicated by 'bad actors' who may choose to disregard UN imperatives in cases such as hacking and cyber-warfare. Despite this complication, and considering the benefits of equitable access, successful international collaboration may help to mitigate the effects of cyber disturbances on society. Institutional Function 5: International collaboration on data, compute, and talent

While collaboration on AI for SDGs is commendable, concerns regarding data privacy, security, and sovereignty must be addressed. Ensuring responsible data sharing, safeguarding intellectual property rights, and respecting cultural and legal differences are essential for fostering trust and cooperation among nations. Regarding the sharing of open-source models, mechanisms should be established to ensure that open-source data is not misused by some players.

Beyond the focus on the right to privacy, other issues should be considered, including nondiscrimination, freedom from xenophobia, etc.

**Institutional Function 6: Reporting and peer review**

Establishing reporting frameworks and mechanisms for peer review are critical for transparency and accountability. Nevertheless, ensuring the integrity and independence of review processes, as well as addressing potential conflicts of interest and biases, are essential to maintain credibility and effectiveness. This is another opportunity to reflect on global disparities that impact access to diverse talent and intellectually qualified peer reviewers to establish a fair pool of individuals to participate in a peer review process.

**Institutional Function 7: Norm elaboration, compliance, and accountability**

We believe that it is important to balance legally binding and non-binding norms to ensure comprehensive AI governance. Balancing the need for enforceability with respect for national sovereignty and cultural diversity requires careful negotiation and compromise. Effective mechanisms for monitoring compliance and addressing violations are essential to ensure accountability and uphold the rule of law.

In addition to the functions outlined in the report, there is arguably a need for a foundational function that addresses issues such as sharing data, public reporting, monitoring, and evidence that AI is working as intended. To ensure that stakeholders have the knowledge and skills to discuss future directions of AI, this foundational function is vital in education and capacity building not just on AI basics, but also legislation for AI and how to adapt to technological advances. We see a role for academia in supporting upskilling, reskilling, continuous education for stakeholders, critical analysis, assessment of regulations, testing and evaluation, and offering perspectives free from financial entanglements from big technology corporations.

Establishing effective collaboration between academia, industry, and government requires clear policies and incentives to promote knowledge sharing, research collaboration, and technology transfer. We have presented a diverse set of examples in our response, which also reflects the multiplicity of scientific and academic disciplines of the authors, and we encourage such interdisciplinary thinking going forward. Ultimately, achieving effective international governance of AI requires a concerted effort from all stakeholders, including governments, intergovernmental organisations, civil society, academia, the private sector, and end users to address complex challenges and ensure AI's benefits are realised while minimising its risks.

## **Any other comments**

The timeline for providing feedback on the report was very short. Larger groups and networks like Responsible AI UK benefit from having more time for in-depth consultations with additional members.

# Response Contributors

(All the authors contributed equally, listed below alphabetically by their first name)

**Professor Bernd Stahl**, Professor of Critical Research in Technology, University of Nottingham.

**Dr. Beverley Townsend**, Research Associate in Law, Ethics, and Technology in the Trustworthy Autonomous Systems (TAS) Resilience Node, University of York.

**Professor Carsten Maple**, Professor of Cyber Systems Engineering, University of Warwick.

**Dr. Charles Vincent**, Reader in AI for Business and Management Science, Queen's Business School, Queen's University Belfast.

**Professor Fraser Sampson**, Professor of Governance & National Security, Centre of Excellence in Terrorism, Resilience, Intelligence & Organised Crime Research, Sheffield Hallam University.

**Professor Geoff Gilbert**, School of Law & Human Rights Centre, University of Essex, Colchester.

**Dr. Helen Smith**, Senior Research Associate in Engineering Ethics for the Trustworthy Autonomous Systems (TAS) Node in Functionality, University of Bristol.

**Jayati Deshmukh**, Senior Research Assistant in Responsible AI, School of Electronics and Computer Science, University of Southampton.

**Dr. Jen Ross**, Senior Lecturer, Centre for Research in Digital Education, University of Edinburgh.

**Dr. Jennifer Williams**, Assistant Professor in Electronics and Computer Science at the University of Southampton.

**Dr. Jesus Martinez del Rincon**, Senior Lecturer in Computer Science, Queen's University Belfast.

**Dr. Justyna Lisinska**, Policy Research Fellow, The Policy Institute, King's College London.

**Dr. Karen O'Shea**, Senior Lecturer (Data Science), School of Engineering and Computing, University of Central Lancashire.

**Dr. Márjory Da Costa Abreu**, Senior Lecturer in Ethical Artificial Intelligence, Sheffield Hallam University.

**Dr. Nelly Bencomo**, Associate Professor in Computer Science, Durham University.

**Oishi Deb**, ELLIS PhD Researcher, MPLS's ED&I Fellow, Visual Geometry Group (VGG) and Torr Vision Group (TVG), Department of Engineering Science and Department of Computer Science, University of Oxford.

**Dr. Peter Winter**, Research Associate in Regulation of Autonomous Systems, Functionality Node, School of Sociology, Politics and International Studies (SPAIS), University of Bristol.

**Dr. Phoebe Li**, Reader in Law and Technology, School of Law, Politics and Sociology, University of Sussex.

**Professor Philip H. S. Torr**, FREng, FRS, Five AI/Royal Academy of Engineering Research Chair in Computer Vision and Machine Learning, Director, International Multimodal Communication Centre (IMCC), Department of Engineering Science, University of Oxford.

**Dr. Pin Lean Lau**, Senior Lecturer (Associate Professor) in Bio-Law, Brunel University London.

**Dr. Raquel Iniesta**, Senior Lecturer (Associate Professor) in Statistical Learning for Precision Medicine, Biostatistics and Health Informatics Department, King's College London.

**Dr. Sarah Kiden**, Senior Research Assistant (Citizen-Centric AI Systems), School of Electronics and Computer Science, University of Southampton.

**Professor Sarvapali D. Ramchurn**, Professor of AI and Director of the UKRI Trustworthy Autonomous Systems (TAS) Hub.

**Professor Sebastian Stein**, Professor of AI and Turing AI Fellow, School of Electronics and Computer Science, University of Southampton.

**Dr. Vahid Yazdanpanah**, Assistant Professor in AI and Computer Science, School of Electronics and Computer Science, University of Southampton.

# Contact details:

For more information about this response, please contact:

- Dr. Sarah Kiden [sk3r24@soton.ac.uk](mailto:sk3r24@soton.ac.uk)
- Professor Sebastian Stein [ss2@ecs.soton.ac.uk](mailto:ss2@ecs.soton.ac.uk) 19
- Professor Sarvapali D. Ramchurn [sdr1@soton.ac.uk](mailto:sdr1@soton.ac.uk)

# About RAI UK

Funded by the Technology Missions Fund, Responsible AI (RAI) UK convenes researchers, industry professionals, policy makers and civil society organisations from across the four nations of the UK to understand how we should shape the development of AI to benefit people, communities, and society. It is an open, multidisciplinary network, drawing on a wide range of academic disciplines. This stems from our conviction that developing responsible AI will require as much focus on the human, and human societies, as it does on AI.